\documentclass
[aps,twocolumn,pre,10pt,superscriptaddress,nobibnotes,showpacs]{revtex4}%
\usepackage{amsfonts}
\usepackage{amsmath}
\usepackage{amssymb}
\usepackage{graphicx}%
\begin{document}
\title{Generation of wakefields by whistlers in spin quantum magnetoplasmas}
\author{A. P. Misra}
\email{apmisra@visva-bharati.ac.in}
\altaffiliation{On leave from Department of Mathematics, Siksha Bhavana, Visva-Bharati University, Santiniketan-731 235, India.}
\affiliation{Department of Physics, Ume{\aa } University, SE--901 87 Ume{\aa }, Sweden.}
\author{G. Brodin}
\email{gert.brodin@physics.umu.se}
\affiliation{Department of Physics, Ume{\aa } University, SE--901 87 Ume{\aa }, Sweden.}
\author{M. Marklund}
\email{mattias.marklund@physics.umu.se}
\affiliation{Department of Physics, Ume{\aa } University, SE--901 87 Ume{\aa }, Sweden.}
\author{P. K. Shukla}
\email{ps@tp4.rub.de; profshukla@yahoo.com}
\affiliation{Department of Physics, Ume{\aa } University, SE--901 87 Ume{\aa }, Sweden.}
\affiliation{RUB International Chair, International Centre for Advanced Studies in Physical
Sciences, Faculty of Physics \& Astronomy, Ruhr University Bochum, D-44780
Bochum, Germany.}


\begin{abstract}
The excitation of electrostatic wakefields in a
magnetized spin quantum plasma by the classical as well as
the spin-induced ponderomotive force (CPF and SPF, respectively) due to whistler waves is reported. The nonlinear dynamics of the whistlers and
the wakefields is shown to be governed by a coupled set of nonlinear
Schr\"{o}dinger (NLS) and driven Boussinesq-like equations. It is found that
the quantum force associated with the Bohm potential introduces two
characteristic length scales, which lead to the excitation of multiple
wakefields in a strongly magnetized dense plasma (with a typical magnetic field
strength $B_{0}\gtrsim10^{9}$ T and particle density $n_{0}\gtrsim10^{36}$
m$^{-3}$), where the SPF strongly dominates over the CPF. In other regimes,
namely $B_{0}\lesssim10^{8}$ T and $\ n_{0}\lesssim10^{35}$ m$^{-3}$, where
the SPF is comparable to the CPF, a plasma wakefield can also be excited
self-consistently with one characteristic length scale. Numerical results
reveal that the wakefield amplitude is enhanced by the quantum tunneling
effect, however it is lowered by the external magnetic field. Under
appropriate conditions, the wakefields can maintain high coherence over
multiple plasma wavelengths and thereby accelerate electrons to extremely high
energies. The results could be useful for particle acceleration at short scales, i.e. at nano- and micrometer scales,
in magnetized dense plasmas where the driver is the whistler wave instead of a laser
or a particle beam.

\end{abstract}
\date{29 November, 2010}
\startpage{1}
\endpage{102}
\maketitle

\section{Introduction}

Lately, significant progress has been made in the field of collective
plasma accelerators for attaining high electron energies
\cite{ProtonDrivenWake,MultipleWake,Wake-E-P-I1,Wake1,MagnetowaveWake,Wake-E-P-I2,Wake2,Wake3,Wake4,WakeMagnetized,WakeLaser}. As is wellknown, an intense electromagnetic (EM) pulse can create a wake of
plasma oscillations under the action of ponderomotive force \cite{WakeLaser}.
Electrons are then trapped into the wake and can thus be accelerated to extremely
high energies, as high as giga- to teraelectronvolt scales. The idea that was presented by Tajima and Dawson
\cite{WakeLaser} three decades ago, has now become a reality through their experimental verifications \cite{RamanInstability,BeatWave}. 
Using intense laser
beams or bunches of relativistic electrons have been used to excite plasma waves
and to produce high electric field strengths (10-100 GV/m), opening
up the possibility for compact GeV particle accelerators
(see, e.g., \cite{ProtonDrivenWake} for a discussion). Recently, an alternative 
scheme has been proposed for accelerating electrons up to the TeV regime, using proton bunches for driving
plasma-wakefield accelerators \cite{ProtonDrivenWake}.

A question that naturally arises is whether or not there exist natural (high-energy) systems, such as astrophysical environments, in which wakefield acceleration can take place.
It is certainly true that neither high-intensity lasers, ultra-short photon bunches, or particle beams, used to excite wakefields in the laboratory regime, are readily available in the astrophysical settings, especially in dense plasmas characterizing, e.g., the interior of
white dwarfs and Jovian planets. In contrast to the above examples of typical drivers, that may
exist independent of a plasma medium, EM whistler waves only exist in a plasma environment.
In this manner, Chen et al.\ \cite{Magnetowave2002} demonstrated the possibility of
exciting large amplitude plasma wakefields by plasma magneto-waves, abundant in astrophysical settings. Later, particle-in-cell simulation has
shown that this mechanism could well be valid for celestial acceleration
\cite{MagnetowaveWake}. However, no attempt has been made to generate
wakefields self-consistently by whistler waves in strongly magnetized dense
quantum plasmas with or without intrinsic spin of electrons.

On the other hand, dense electron plasmas are degenerate, and one must
therefore take into account Fermi-Dirac statistics as well as
tunneling effects \cite{PhysUsp}. Furthermore, apart from the statistical effects introduced by the spin-$1/2$ nature of the electrons, the spin also generates a ``dynamical" contribution, in the form of a pressure-like spin force and a
spin magnetization current \cite{Spin-Nat,Spin}. The latter can even be larger
than the classical free current when low-frequency (lf) longitudinal
perturbations are driven by a spin-induced ponderomotive nonlinearity in the
propagation of a short EM pulse \cite{SPF}. Naturally, the inclusion of the
quantum statistical pressure, the quantum force associated with the Bohm
potential as well as the spin force due to finite magnetic moment of electrons
in the dynamical equations, will significantly change the collective behavior of the 
electrons. For example, the Bohm potential, arising due to the particle's
wave-like nature, provides higher-order dispersion as well as the possibility
of short-wavelength (compared to the electron skin depth) wakefield. The spin
magnetization contributes to linear dispersion of circularly polarized modes
\cite{JPPMisra} and the magnetic dipole force results in a spin contribution
to the ponderomotive force \cite{SPF}.

In this paper, we present a theoretical investigation of the excitation of
multiple wakefields driven by both the classical (CPF) \cite{CPF} as well as the
spin ponderomotive force (SPF) \cite{SPF} of EM whistlers in a magnetized
quantum plasma with intrinsic magnetization. The self-consistent fields of the 
whistlers and the wake are described by a coupled set of nonlinear
Schr\"{o}dinger- and modified Boussinesq-like equations \cite{WhistlerSpin}. It
is shown that when the SPF is comparable to CPF, the wakefields can be
generated in the regime of strong magnetic fields with $B_{0}\lesssim10^{8}$T
and high-density plasmas with $\ n_{0}\lesssim10^{35}$m$^{-3}.$ However,
multiple wakefields are also excited when the SPF dominates over the
CPF in the very strongly magnetized and superdense plasmas with $B_{0}%
\gtrsim10^{9}$T, $n_{0}\gtrsim10^{36}$m$^{-3}.$ In the latter, the results of our model is that the whistler
wavelength may be even shorter than the Compton wavelength of the electrons. This
gives rise to a viable mechanism for wakefield acceleration
in spin dominated plasmas, such as those in superdense
astrophysical bodies, e.g., white dwarfs, neutron stars, magnetars.

\section{Evolution equations}

We consider the propagation of high-frequency (hf) whistler waves along an
external magnetic field $\mathbf{B}=B_{0}\hat{z}$ in a quantum plasma, taking into account the 
intrinsic spin of the electrons. The heavy ions are assumed to be immobile, so
that the lf density response occurs on a time-scale much shorter than the ion
plasma period. In the modulational representation, the whistler electric field
can be represented by $\mathbf{E}=\left(  \hat{x}-i\hat{y}\right)
E(z,t)\exp(ikz-i\omega t)+$c.c., where $E(z,t)$ is the slowly varying wave
envelope, $\omega$ ($k$) is the whistler wave frequency (number) and c.c. stands
for the complex conjugate. Then the basic equations for the evolution of
whistlers consist of the fluid equations for the nonrelativistic evolution of
spin$-1/2$ electrons given by \cite{Spin, WhistlerSpin}%

\begin{equation}
\partial_{t}n_{e}+\nabla.\left(  n_{e}\mathbf{v}_{e}\right)  =0 \label{b1}%
\end{equation}%
\begin{align}
\left(  \partial_{t}+\mathbf{v}_{e}.\nabla\right)  \mathbf{v}_{e}  &
=-\frac{e}{m_{e}}\left(  \mathbf{E}+\mathbf{v}_{e}\times\mathbf{B}\right)
-\frac{\nabla P_{e}}{m_{e}n_{e}}\nonumber\\
& \quad  +\frac{\hbar^{2}}{2m_{e}^{2}}\nabla\left(  \frac{\nabla^{2}\sqrt{n_{e}}%
}{\sqrt{n_{e}}}\right)  +\frac{2\mu}{m_{e}\hbar}\mathbf{S}.\nabla B \label{b2}%
\end{align}%
\begin{equation}
\left(  \partial_{t}+\mathbf{v}_{e}.\nabla\right)  \mathbf{S}=-\left(
2\mu/\hbar\right)  \left(  \mathbf{B}\times\mathbf{S}\right)  , \label{b3}%
\end{equation}
and the Maxwell equations for the electromagnetic (EM) fields
\begin{equation}
\nabla\times\mathbf{E}=-\partial_{t}\mathbf{B},\text{ }\nabla.\mathbf{B}=0
\label{b4}%
\end{equation}%
\begin{equation}
\nabla\times\mathbf{B}=\mu_{0}\left(  \varepsilon_{0}\frac{\partial\mathbf{E}%
}{\partial t}\mathbf{-}en_{e}\mathbf{v}_{e}\mathbf{+}\frac{2\mu}{\hbar}%
\nabla\times n_{e}\mathbf{S}\right)  . \label{b5}%
\end{equation}
Here $n_{e},$ $m_{e},$ $\mathbf{v}_{e}$ denote the number density, mass and
velocity of electrons respectively, $\mathbf{B}$ is the magnetic field and
$P_{e}$ is the electron thermal pressure and $\mathbf{S}$ is the spin angular
momentum with $\left\vert \mathbf{S}\right\vert =\left\vert S_{0}\right\vert
\equiv\hbar/2,$ $\mu=-\left(  g/2\right)  \mu_{B},$where $g\approx2.0023193$
is the electron $g$-factor and $\mu_{B}\equiv e\hbar/2m_{e}$ is the Bohr
magneton. The equations (\ref{b1})-(\ref{b3}) are applicable even when
different spin states (with up and down) are well represented by a macroscopic
average. This may, however, occur in the regimes of very strong magnetic
fields (or a very low temperature plasmas), where generally the electrons
occupy the lowest energy spin states. On the other hand, for a time-scale
larger than the spin-flip frequency, the macroscopic spin state (to be
attenuated by a factor decreasing the effective value of $\left\vert
\mathbf{S}\right\vert $ below $\hbar/2$) can \ be well-described by the
thermodynamic equilibrium spin configuration, and in this case the above fluid
model can still be applied. However, this is not the present issue to be
studied here rather we will focus on the regimes of strong magnetic fields and
high density plasmas.

The evolution equation for the whistler can be obtained by taking the curl of Eq. (\ref{b2}) [hence the pressure gradient and the quantum force  in Eq. (\ref{b2}) vanish] and using Eqs. (\ref{b3})-(\ref{b5}) as %
\begin{widetext}
\begin{align}
0& =\frac{e}{m_{e}}\frac{\partial \mathbf{B}}{\partial t}+\frac{\varepsilon
_{0}}{en_{e}}\frac{\partial }{\partial t}\left( \frac{\partial ^{2}\mathbf{B}%
}{\partial t^{2}}+\frac{1}{n_{e}}\nabla n_{e}\times \frac{\partial \mathbf{E}%
}{\partial t}\right) -v_{ez}\nabla \times \frac{\partial \mathbf{v}_{e}}{%
\partial z}+\frac{1}{e\mu _{0}}\frac{\partial }{\partial t}\left[ \frac{1}{%
n_{e}}\nabla \times \left( \nabla \times \mathbf{B}\right) \right]   \notag
\\
& -\frac{2\mu }{e\hbar }\frac{\partial }{\partial t}\left[ \frac{1}{n_{e}}%
\nabla \times \left( \nabla \times n_{e}\mathbf{S}\right) \right] +\frac{1}{%
m_{e}\mu _{0}n_{e}}\nabla \times \left[ \left( \nabla \times \mathbf{B}%
\right) \times \mathbf{B}\right] +\frac{2\mu }{m_{e}\hbar }\nabla \times
\left( S^{a}\nabla B_{a}\right)   \notag \\
& -\frac{\varepsilon _{0}}{m_{e}n_{e}}\nabla \times \left( \frac{\partial
\mathbf{E}}{\partial t}\times \mathbf{B}\right) -\frac{2\mu }{m_{e}\hbar
n_{e}}\nabla \times \left[ \left( \nabla \times n_{e}\mathbf{S}\right)
\times \mathbf{B}\right].  \label{ev}
\end{align}%
\end{widetext}

We note that in  the nonlinear interaction of hf EM waves with the lf electron plasma response, the use of cold plasma approximation is also justified to the
 fact that for large field intensities and moderate electron temperature, the directed
speed of electrons in the hf fields is much larger than the random thermal speed.  It can also be shown that the density perturbation
 associated with the hf EM wave is zero. Thus, the evolution equation (\ref{ev}) for the whistlers do not involve contributions from the electron pressure and the quantum tunneling effect proportional to $\hbar^2$. Later, we will see that these contributions will appear in the coupling of the lf plasma response with the hf one through the ponderomotive force induced by the hf field.
 
Introducing the variables $B_{\pm}=B_{x}\pm iB_{y},$ $E_{\pm}=E_{x}\pm iE_{y}$
etc., suitable for circularly polarized waves, and linearizing we obtain respectively from the Faraday's law  and the spin-evolution
 equation (\ref{3}) as \cite{SPF}%
\begin{equation}
B_{\pm}=\pm\frac{ik}{\omega}E_{\pm},\text{ }S_{\pm}=\mp\frac{2\mu\left\vert
S_{0}\right\vert B_{\pm}}{\hbar\left(  \omega\pm\omega_{g}\right)  } \label{bs}%
\end{equation}
We then linearize  Eq. (\ref{ev})  and use Eq. (\ref{bs}) to obtain the following
linear dispersion relation  for the circularly polarized modes  \cite{JPPMisra}
\begin{equation}
n_{R}^{2}=1-\frac{\omega_{pe}^{2}}{\omega\left(  \omega\pm\omega_{c}\right)
}-\frac{g^{2}\omega_{pe}^{2}k^{2}\left\vert S_{0}\right\vert }{4\omega
^{2}m_{e}\left(  \omega\pm\omega_{g}\right)  }, \label{dis}%
\end{equation}
which can be rewritten as
\begin{equation}
n_{R}^{2}\left(  1+\frac{\omega_{\mu}}{\omega-\omega_{g}}\right)
=1-\frac{\omega_{pe}^{2}}{\omega\left(  \omega-\omega_{c}\right)  }, \label{1}%
\end{equation}
where $n_{R}\equiv ck/\omega$ is the refractive index, $\omega_{\mu}%
=g^{2}\left\vert S_{0}\right\vert /4m_{e}\lambda_{e}^{2}$ is the frequency due
to the plasma magnetization current and $\lambda_{e}\equiv c/\omega_{pe}$ is
the electron skin depth with $\omega_{pe}\equiv\sqrt{n_{0}e^{2}/\varepsilon
_{0}m_{e}}$ denoting the electron plasma frequency. Moreover, $\omega
_{c}=eB_{0}/m_{e}\ $is the electron-cyclotron frequency and $\omega
_{g}=(g/2)\omega_{c}$ \ is the electron spin-precession frequency. We note
that the frequency resonances occur not only at the cyclotron frequency
($\omega\rightarrow\omega_{c}$) but also due to the spin-gyration of electrons
($\omega\rightarrow\omega_{g}$), although these resonances are close to each
other. At the resonance, the transverse field associated with the whistlers
rotates at the same speed as electrons gyrates around the external magnetic
field. The electrons will then experience a continuous acceleration from the
wave electric field. However, the detail discussion on the properties of the
linear modes modified by the spin magnetization current can be found in the
literature \cite{JPPMisra}. One can see that when $\omega_{c}\gg\omega_{pe},$
i.e., the magnetic field is strong enough, but $\omega_{\mu}$ is still smaller
than $\omega_{pe}$ (unless we consider a very high-density regime), the
dispersion of the whistler wave are almost linear (with phase speed
approaching the speed $c$ of light in vacuum$)$ over a wide range of wave
numbers. However, it may remain nonlinear in strongly magnetized high density
plasmas depending on the parameter regimes we consider \cite{JPPMisra}. This
nonlinear behaviors of whistlers in which $E,$ $B$ wave fields are not
comparable in strength do not favor the excitation of wakefields, and we do
not consider those cases here.\

On the other hand, in the nonlinear regime, the dynamics of whistler wave
envelopes can be described from  Eq. (\ref{ev}) by the following nonlinear Schr\"{o}dinger
(NLS)-like equation \cite{WhistlerSpin}.
\begin{equation}
i\left(  \partial_{t}E+v_{g}\partial_{z}E\right)  +\left(  v_{g}^{\prime
}/2\right)  \partial_{z}^{2}E-\Delta E=0, \label{2}%
\end{equation}
where $E\equiv E_{x}-iE_{y},$ $v_{g}\equiv d\omega/dk$ \ is the group speed,
 $v_{g}^{\prime}\equiv d^{2}\omega/dk^{2}$ is the group dispersion of
whistlers and $\Delta$ is the nonlinear frequency shift. These are given by \cite{WhistlerSpin} 
\begin{widetext}
\begin{equation}
v_{g}=\left(  \frac{2c^{2}k}{\omega_{pe}^{2}}+\frac{g^{2}\hbar k}%
{4m_{e}\left(  \omega-\omega_{g}\right)  }\right)  /\left(  \frac{2\omega
}{\omega_{pe}^{2}}+\frac{\omega_c}{\left(  \omega-\omega_c\right)  ^{2}}%
+\frac{g^{2}\hbar k^{2}}{8m_{e}\left(  \omega-\omega_{g}\right)  ^{2}}\right)
, \label{vg1}%
\end{equation}%
\begin{equation}
v_{g}^{\prime}=\frac{v_{g}}{k}\left[  1-\frac{2kv_{g}^{2}}{\Lambda\omega
_{pe}^{2}}\left(  1-\frac{\omega_c\omega_{pe}^{2}}{\left(  \omega-\omega_c\right)
^{3}}\right)  -\frac{g^{2}\hbar k^{2}v_{g}}{4m_{e}\Lambda\left(  \omega
-\omega_{g}\right)  ^{2}}\left(  2-\frac{v_{g}k}{\omega-\omega_{g}}\right)
\right],\label{vg2}
\end{equation}
\end{widetext} and
\begin{equation}
\Delta=\frac{v_{g}}{\Lambda}\left[  \frac{k\omega v_{ez}}{\left(
\omega-\omega_{c}\right)  ^{2}}+\left(  \frac{\omega}{\omega-\omega_{c}}%
+\frac{g^{2}\hbar k^{2}}{4m_{e}\left(  \omega-\omega_{g}\right)  }\right)
N\right], \label{3}%
\end{equation}
where $\Lambda=2c^{2}k/\omega_{pe}^{2}+g^{2}\hbar k/4m_{e}\left(
\omega-\omega_{g}\right)  ,$ $N\equiv n_{e}/n_{0}$ and $v_{ez}$ is the
magnetic field aligned free electron flow speed. 

Note that the variables $v_{ez}$ and $N$  appear due to  the lf  electron plasma response, and are to be coupled with  hf field  through the ponderomotive force. The governing equations for the nonlinear coupling of these two responses  will be described later.  Furthermore, in Eqs. (\ref{vg1})-(\ref{3}), the terms proportional to $\hbar$ are contributions from the plasma
magnetization current due to intrinsic spin of electrons.   Notice, however, that when the spin
effects dominate, the group velocity tends to decrease as the frequency of the
whistler approaches the cyclotron frequency \ \cite{JPPMisra} and/ or $k$
increases its value. The latter indicates a negative group dispersion in the
propagation of whistlers. Notice, however, that the nonlinear frequency shift
precisely depends on the group velocity, and can even become larger (when
$\omega$ approaches $\omega_{c})$ due to the plasma streaming with the flow
speed $v_{ez}$ along the external magnetic field.

Let us now compare the two ponderomotive forces, which induce slowly varying
electrostatic oscillations in plasmas. The expressions of these forces, namely
CPF \cite{CPF} and SPF \cite{SPF} [Note that in the expression for the spin ponderomotive force given by Eqs. (12) and (13) in Ref. \cite{SPF}, 
a factor $2$ was missing in the denominator, and this factor $2$  appears in the average of the forces as in Eqs. (7) and (8) there for the classical part] acting on an individual electron can be
written respectively as%

\begin{equation}
F_{cz}=-\frac{1}{2}\frac{e^{2}}{m_{e}^{2}\omega\left(  \omega-\omega
_{c}\right)  }\left(  1+\frac{\omega_{c}}{\omega}\frac{kv_{g}}{\omega
-\omega_{c}}\right)  \frac{\partial\left\vert E\right\vert ^{2}}{\partial\xi},
\label{CPF}%
\end{equation}%
\begin{equation}
F_{sz}=-\frac{1}{4}\left(  \frac{g^{2}\hbar k^{2}}{4m_{e}\omega}\right)
\frac{e^{2}}{m_{e}^{2}\omega\left(  \omega-\omega_{g}\right)  }\left(
1+\frac{kv_{g}}{\omega-\omega_{g}}\right)  \frac{\partial\left\vert
E\right\vert ^{2}}{\partial\xi}, \label{SPF}%
\end{equation}
where $\xi=z-v_{g}t$ is the co-moving frame of reference for the driving
whistler. The spin induced ponderomotive force arises due to the effects of the finite magnetic moment of electrons and is obtained by taking the average over the fast time scale of the spin force in the momentum equation (\ref{b2}) \cite{SPF}. We note that the spin contribution to the ponderomotive force is
small compared to the classical one when the factor $\chi\equiv$ $\hbar
k^{2}/m_{e}\omega\ll1.$ In this case, one can typically neglect the
spin-contribution in the linear as well as nonlinear regimes, and thus the
results will be valid for weakly or strongly magnetized low-density plasmas.
However, for an exception to this rule, see Ref. \cite{SPF}. In the opposite
limit, i.e., \ $\chi\gg1,$ the SPF in Eq. (\ref{SPF}) can, indeed, dominate
over the CPF when the whistler wave frequency is close to the cyclotron
frequency. In this case, the whistler (with phase speed close to $c)$
wavelength $\left(  \lambda_{W}=2\pi/k\right)  $ can be smaller or of the
order of the Compton wavelength $\left(  \lambda_{C}=h/mc\right)  $ of
electrons, since $\hbar k^{2}/m_{e}\omega\ $scales as $\lambda_{C}/\lambda
_{W}$. This case can be relevant in \ very strongly magnetized ($B_{0}%
\gtrsim10^{9}$T) and very high-density medium ($n_{0}\gtrsim10^{36}$m$^{-3})$
in which $\omega_{c}>\omega_{pe}$ still holds$.$ The whistler wave dispersion
can then still be linear and the group as well the phase speed of the
whistlers may approach $\ $the speed of light in vacuum. Below, we will see
that this is the case in which the whistler pulses can maintain their shape
over a macroscopic distance, and thus favors the plasma wakefield acceleration.

Next, the whistler dispersion may be linear in the regimes of $B_{0}\sim
10^{8}$ T and $n_{0}\sim10^{35}$ m$^{-3}$ where the SPF may not be dominant
over, but may be comparable to the CPF $F_{cz}$. On the other hand, for
$\omega\ll\omega_{c}$ and $\chi\gg1,$ though the first term in the square
brackets in Eq. (\ref{CPF}) is negligible, but the second one can not be so,
rather may be comparable to the spin contribution. In this case, ion dynamics
might play roles, and also the whistler wavelength may be much larger than the
plasma characteristic length scale. So, this case will not be so effective for
high energy particle acceleration. We will focus on the regimes as discussed
above for which the excitation of multiple wakefields is possible. For
$\omega\gg\omega_{c}$ and $\chi\gg1,$ the classical part is also negligible,
however, this is not relevant \ to the present study.

The equations for the lf density response which arises due to the ponderomotive force of the hf electromagnetic field satisfy the electron continuity, momentum balance and the Poisson
equation, these are respectively%

\begin{equation}
\partial_{t}N+\partial_{z}v_{ez}=0. \label{4}%
\end{equation}%
\begin{align}
&  \partial_{t}v_{ez}+\frac{e}{m_{e}}E_{l}+V_{F}^{2}\partial_{z}N-\left(
\hbar^{2}/4m_{e}^{2}\right)  \partial_{z}^{3}N\nonumber\\
&  =\frac{e^{2}}{2m_{e}^{2}\omega^{2}}\left(  \Gamma_{1}\partial_{z}%
|E|^{2}-k\Gamma_{2}\partial_{t}|E|^{2}\right)  , \label{5}%
\end{align}%
\begin{equation}
\partial_{z}E_{l}=\frac{e}{\varepsilon_{0}}(n_{0}-n_{e}), \label{9}%
\end{equation}
where $E_{l}$ is the lf part of the wave electric field (wakefield) and
$V_{F}$ $=\sqrt{k_{B}T_{F}/m_{e}}$ \ is the Fermi speed relevant for a
low-temperature high density plasmas \cite{FermiPressure}, $T_{F}\equiv
\hbar^{2}\left(  3\pi^{2}n_{0}\right)  ^{2/3}/2k_{B}m_{e}$ is the Fermi
temperature with $k_{B}$ denoting the Boltzmann constant. The term
$\propto\hbar^{2}$ is the effect of quantum tunneling associated with the Bohm
de Broglie potential. The ponderomotive force contributions are proportional
to the constants $\Gamma_{1}$ and $\Gamma_{2}$ \ \cite{WhistlerSpin} where%
\begin{align}
\Gamma_{1}  &  =\frac{\omega}{\omega-\omega_{c}}+\frac{g^{2}\hbar k^{2}%
}{8m_{e}\left(  \omega-\omega_{g}\right)  },\nonumber\\
\Gamma_{2}  &  =\frac{\omega_{c}}{\left(  \omega-\omega_{c}\right)  ^{2}%
}+\frac{g^{2}\hbar k^{2}}{8m_{e}\left(  \omega-\omega_{g}\right)  ^{2}},
\label{6}%
\end{align}
in which the first terms appear due to CPF \cite{CPF} and the second $\left(
\propto\hbar\right)  $ are due to the SPF \cite{SPF}. \ From Eq. (\ref{6}) we
note that for $\chi\gg1,$ the SPF contribution is substantial compared to the
CPF$,$ and $\Gamma_{1,2}$ can be approximated as $\propto\hbar$ that
represents purely a spin quantum effect. Equations (\ref{4})-(\ref{9}) can be
combined to obtain the driven wave equation for lf perturbations of the
Boussinesq-type
\begin{equation}
\left[  \partial_{t}^{2}+\frac{\hbar^{2}}{4m_{e}^{2}}\partial_{z}^{4}%
-V_{F}^{2}\partial_{z}^{2}+\omega_{pe}^{2}\right]  N=\mu_{1}\partial_{z}%
^{2}|E|^{2}-\mu_{2}\partial_{zt}^{2}|E|^{2}, \label{10}%
\end{equation}
where the ponderomotive force contributions are $\mu_{1}=\varepsilon_{0}%
\omega_{pe}^{2}\Gamma_{1}/2m_{e}\omega^{2}\ $and $\mu_{2}=\varepsilon
_{0}\omega_{pe}^{2}k\Gamma_{2}/2m_{e}\omega^{2}$ in which spin effect is hidden.

Thus, we have a new set of three coupled equations, namely (\ref{2}),
(\ref{4}) and (\ref{10}), modified by the SPF and the effect of quantum
tunneling, which describes the nonlinear coupling of electron whistlers with
the field aligned electrostatic density fluctuations. These equations can be
written in the nondimensional forms as ($z\rightarrow z/\lambda_{Fe}%
,t\rightarrow t\omega_{pe},E\rightarrow E/E_{0},$ $v_{ez}\rightarrow
v_{ez}/c_{s}$, where $\lambda_{Fe}=V_{F}/\omega_{pe}=\lambda_{e}V_{F}/c$)
\begin{equation}
i\left(  \partial_{t}E+V_{g}\partial_{z}E\right)  +\left(  V_{g}^{\prime
}/2\right)  \partial_{z}^{2}E-\Psi E=0, \label{11}%
\end{equation}%
\begin{equation}
\partial_{t}N+\partial_{z}v_{ez}=0, \label{12}%
\end{equation}%
\begin{equation}
\left[  \partial_{t}^{2}+H^{2}\partial_{z}^{4}-\partial_{z}^{2}+1\right]
N=\zeta_{1}\partial_{z}^{2}|E|^{2}-\zeta_{2}\partial_{zt}^{2}|E|^{2},
\label{13}%
\end{equation}
where $E_{0}=\sqrt{2k_{B}T_{F}n_{0}/\varepsilon_{0}},$ $V_{g}=v_{g}%
/V_{F},V_{g}^{\prime}=v_{g}^{\prime}\omega_{pe}/V_{T}^{2},$ $\Psi
=\Delta/\omega_{pe},$ $H=\hbar\omega_{pe}/2k_{B}T_{F}$ \ is the quantum
parameter and $\zeta_{1}=\omega_{pe}^{2}\Gamma_{1}/\omega^{2},$ $\zeta
_{2}=\omega_{pe}^{2}kV_{F}\Gamma_{2}/\omega^{2}.$ Equations (\ref{11}%
)-(\ref{13}) contain the main results of the present work.

\section{Excitation of the wakefields}

In this section, we consider the excitation of one-dimensional wakefield by
the hf whistlers propagating with the group velocity. Thus, we look for
stationary solutions of\ our main Eqs. (\ref{11})- (\ref{13}) in the frame
$\xi=z-V_{g}t$.\ To this end, we assume $E$ to be of the form $E=W\left(
\xi\right)  \exp\left(  -i\Theta t\right)  ,$ where $\ W$ is a real function
normalized by $E_{0}$ and $\Theta$ is a real constant. Then Eqs. (\ref{11})-
(\ref{13}) reduce to
\begin{equation}
\left(  V_{g}^{\prime}/2\right)  d_{\xi}^{2}W+W\Omega-\tilde{\Delta
}NW=0,\label{18}%
\end{equation}%
\begin{equation}
\left[  H^{2}d_{\xi}^{4}+\left(  V_{g}^{2}-1\right)  d_{\xi}^{2}+1\right]
N=\left(  \zeta_{1}+\zeta_{2}V_{g}\right)  d_{\xi}^{2}W^{2},\label{19}%
\end{equation}
where
\begin{equation}
\tilde{\Delta}=\frac{v_{g}}{\Lambda\omega_{pe}}\frac{\omega}{\omega-\omega
_{c}}\left(  1+\frac{kv_{g}}{\omega-\omega_{c}}+\frac{g^{2}\hbar k^{2}}%
{4m_{e}\omega}\frac{\omega-\omega_{c}}{\omega-\omega_{g}}\right)  .\label{20}%
\end{equation}
Equation (\ref{19}) shows that the electrostatic wake field is created by the
self-consistent whistler waves$.$ \ We also find that the inclusion of the
quantum correction $\propto H$ leads to\ two different characteristic length
scales, which, in turn, will give rise multiple oscillatory wakefields. Thus,
considering a trial solution of the corresponding homogeneous equation of
(\ref{19}) as $N\sim\exp(ik_{p}\xi),$ we obtain $k_{p}^{4}-\kappa_{a}^{2}%
k_{p}^{2}+\kappa_{b}^{4}=0,$ where $\kappa_{a}=(V_{g}^{2}-1)^{1/2}/H\ $\ and
$\ \kappa_{b}=1/\sqrt{H}.$ This biquadratic equation has, in general, two
pairs of real roots $\pm k_{1,2}$ \cite{MultipleWake} for $V_{g}\equiv
v_{g}/V_{F}>\sqrt{1+2H}$ where
\begin{equation}
k_{1,2}^{2}=\frac{\kappa_{a}^{2}}{2}\mp\sqrt{\frac{\kappa_{a}^{4}}{4}%
-\kappa_{b}^{4}}.
\end{equation}
For a Gaussian driving pulse of the form $W\sim\exp\left(  -\xi^{2}/L_{p}%
^{2}\right)  ,$ Eq. (\ref{19}) yields the wakefield behind the driving pulse,
i.e., for $\left\vert \xi\right\vert \gg L_{p}$ as $N=N_{2}-N_{1}$ where%

\begin{equation}
N_{1,2}\approx\frac{4\kappa_{b}^{4}\left(  \zeta_{1}+\zeta_{2}V_{g}\right)
}{\left(  12-L_{p}^{2}k_{1,2}^{2}\right)  \left(  k_{2}^{2}-k_{1}^{2}\right)
}\cos\left(  k_{1,2}\xi\right)
\end{equation}
and, in particular, in absence of the quantum force,%

\begin{equation}
N\approx\frac{4\kappa_{p}^{2}\left(  \zeta_{1}+\zeta_{2}V_{g}\right)
}{\left(  12-L_{p}^{2}k_{p}^{2}\right)  }\cos\left(  \kappa_{p}\xi\right)  ,
\end{equation}
where $\kappa_{p}=1/(V_{g}^{2}-1)^{1/2}.$

Next, for the generation of whistler driven wakefields, we numerically solve
the Eqs. (\ref{18}) and (\ref{19}) by Newton's method with the boundary
conditions $N$, $W$, $dN/d\xi,$ $dW/d\xi\rightarrow0$ as $\xi\rightarrow
\pm\infty.$ In order to study the influence of the quantum tunneling effect,
we consider two cases, namely for $H=0$ [Note, however, that $H$ $\left(
\sim1/\hbar\right)  =0$ does not mean that one recovers the $\hbar=0$ case$,$
rather it implies that we have simply disregarded the quantum tunneling effect
associated with the Bohm potential] and for nonzero $H.$ Here the typical
length scale for plasma collective oscillation is the Fermi screening length
$\lambda_{Fe}=\lambda_{e}V_{F}/c<\lambda_{e}$ as long as the $V_{F}<c,$ and it
must be such that $\lambda_{Fe}<\lambda_{W},$ the whistler wavelength$.$ The
typical time scale is the electron plasma period, which is shorter than the
ion plasma period for which the ion dynamics is negligible. \ Moreover,
quantum effects become important when $\zeta\equiv T_{F}/T_{classical}=\left(
1/2\right)  \left(  3\pi^{2}n_{0}\lambda_{B}^{3}\right)  ^{2/3}\gtrsim1,$
where $\lambda_{B}=\hbar/m_{e}V_{F}$ is the thermal de Broglie wavelength.
Furthermore, the quantum coupling parameter $g_{Q}\sim\left(  1/n_{0}%
\lambda_{Fe}^{3}\right)  ^{2/3}\lesssim1$ defines a quantum collisionless
regime where the collective and mean-field effects dominate. We now study the
two cases as follows.

In the case of $H=0$, the plasma characteristic length scale for the wake is
$\lambda_{p}=2\pi/\kappa_{p},$ and we choose the parameters as $B_{0}%
=3\times10^{8}$ T, $n_{0}=10^{35}$ m$^{-3}$ and $\omega=0.78,$ $\Theta=0.22,$
for which $\omega/\omega_{p}=2.3,$ $\omega_{c}/\omega_{p}=3,v_{g}%
/V_{F}=1.3,v_{g}/v_{p}=0.7,$ $\lambda_{Fe}/\lambda_{e}=0.4$ and $E_{0}%
=1.7\times10^{16}$V/m. In this case the plasma oscillation length is
$\lambda_{p}(=5.4)>\lambda_{W}$ $(\equiv2\pi/k\lambda_{Fe}$ in normalized
form$)=5.3,$ the group $\ $dispersion $V_{g}^{\prime}=-2.7$ and the nonlinear
frequency shift $\tilde{\Delta}=0.4.$ Moreover, $g_{Q}=0.11<1$ and $\zeta
\sim1.$ Thus, in the quantum regime, the short-scale wakefield can be
generated as shown in the upper panel of Fig. 1. \ Increasing the magnetic
field strength, namely $B_{0}=6\times10^{8}$ T and keeping electron
concentration as the same, i.e., $n_{0}=10^{35}$ m$^{-3}$, we find that the
group velocity increases ($v_{g}/v_{p}=0.85$) and hence for a increasing value
of the ratio $\omega_{c}/\omega_{p}=5.9,$ the wakefield remains coherent, but
its amplitude becomes much lowered (not shown in the figure). \ The whistler
group dispersion has now been lowered as $V_{g}^{\prime}=-1.5$ than the
previous case and the frequency is shifted-up to $\tilde{\Delta}=0.53$. The
other parameter values in this regime are, e.g., $\lambda_{p}=10.9$ and
$\lambda_{W}=3.2.$ \begin{figure}[ptb]
\begin{center}
\includegraphics[height=3.5in,width=3.5in]{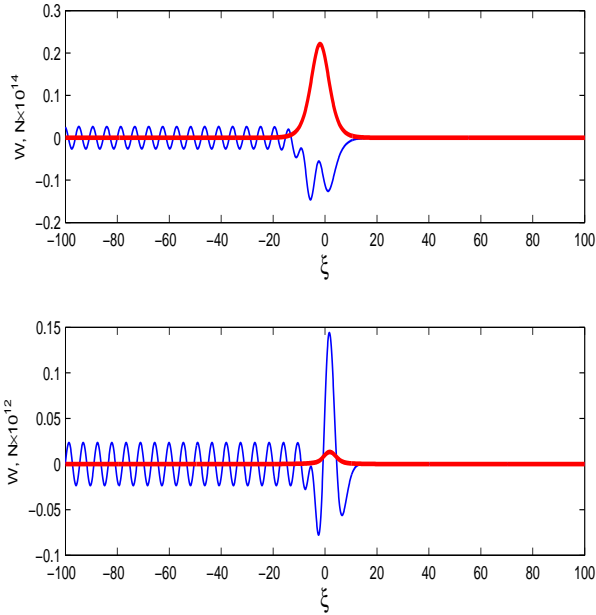}
\end{center}
\caption{(Color online) Plasma wakefields [Numerical solution of Eqs.
(\ref{18}) and (\ref{19})] driven by the whistlers (thick line) for $H=0$
(upper panel) and for $H=0.13$ (lower panel) with characteristic length scales
$\lambda_{p}=5.4$ $(\lambda_{Fe})$ and \ $\lambda_{1}=5.5$ $(\lambda_{Fe})$
respectively. The case in which SPF is comparable to CPF. The parameter values
are given in the text. }%
\end{figure}

The lower panel of Fig. 1 shows the wakefields for a\ nonzero $H$. \ We find
that the plasma wake electric field is enhanced due to the quantum tunneling
effect. \ Here the corresponding plasma characteristic length scales are
$\lambda_{1,2}=2\pi/$ $k_{1,2}$ as defined above. For the length scale
$\lambda_{1},$ we use the parameters $B_{0}=6\times10^{8}$ T, $n_{0}%
=2\times10^{35}$ m$^{-3},$ $\omega=0.78\ $and $\Theta=0.04$ for which
$\omega/\omega_{p}=3.3,$ $\omega_{c}/\omega_{p}=4.2,$ $v_{g}/V_{F}=1.3,$
$v_{g}/v_{p}=0.77,$ $k\lambda_{Fe}=1.9,$ $H=0.13,$ $\lambda_{1}=5.5,$
$\lambda_{W}=3.3,V_{g}^{\prime}=-1.5$ and $\tilde{\Delta}=0.5.$ We note that
the length scale $\lambda_{2}$ does not favor the wakefield generation in the
parameter regimes considered above, since for $V_{g}>\sqrt{1+2H},$
$\lambda_{2}$ scales as the Compton wavelength, which becomes lower than the
whistler wavelength $\lambda_{W}$. \ However, since as said before,
$\lambda_{W}$ may approach or lower than the Compton wavelength in the case of
very strongly magnetized and very high density plasmas where the SPF dominates
over the CPF, multiple wakefield generation can be possible as can be
illustrated from Fig. 2 below. \begin{figure}[ptb]
\begin{center}
\includegraphics[height=3.5in,width=3.5in]{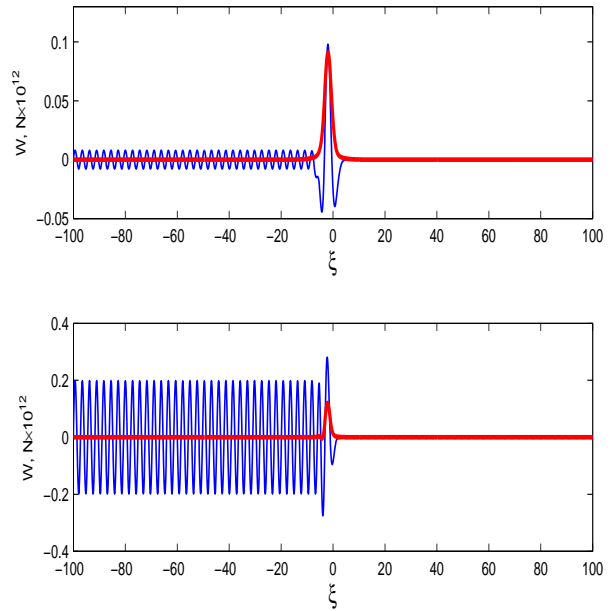}
\end{center}
\caption{(Color online) Plasma wakefields [Numerical solution of Eqs.
(\ref{18}) and (\ref{19})] driven by the whistlers (thick line) for a fixed
$H=0.13$ but for two different characteristic length scales $\lambda_{1}=2.8$
$(\lambda_{Fe})$ (upper panel) and $\lambda_{2}=1.45$ $(\lambda_{Fe})$ (lower
panel). The case in which SPF dominates over CPF. The parameter values are
given in the text. }%
\end{figure}Thus,  for the parameters $B_{0}=4\times10^{9}$ T, $n_{0}=10^{36}$
m$^{-3}$ and $\omega=0.78,$ $\Theta=0.008$ for which $\omega/\omega_{p}=9.7,$
$\omega_{c}/\omega_{p}=12.5,v_{g}/V_{F}=1.1,v_{g}/v_{p}=0.97,$ $k\lambda
_{e}=9.97,$ $k\lambda_{Fe}=8.4,$ $2\pi/k\lambda_{e}=0.63,$ $H=0.13,$
$\lambda_{W}=0.75,V_{g}^{\prime}=-0.1,$ $\tilde{\Delta}=0.3,$ the
corresponding wakefield excitation is shown in the upper panel of Fig. 2 with
$\lambda_{1}=2.8.$ From the lower panel of this figure one can see the similar
excitation but with a shorter scale $\lambda_{2}=1.45.$ In the latter,
wakefield amplitude is seen to be enhanced compared to the case with
$\lambda_{1}.$

\section{Discussion and conclusion}

Numerical results in the previous section suggest that wakefield excitation by
the whistlers can be possible in a strongly magnetized dense plasmas (
$B_{0}\lesssim10^{8}$ T and $\ n_{0}\lesssim10^{35}$ m$^{-3})$ with intrinsic
spin of electrons. In this regime the SPF is no longer negligible rather may
be comparable to the CPF. Such parameter regimes can be achievable in the
interior of white dwarf stars where electrons are indeed degenerate and
the\ electron degeneracy pressure is what supports a white dwarf against
gravitational collapse. The magnetic fields in such compact degenerate stars
might be due to conservation of total surface magnetic flux during the
evolution or by the \textbf{\ }emission of circularly polarized light or
else\textbf{. }However, a small number of white dwarfs have been examined for
magnetic field strength exceeding $10^{5}$ T. \

In absence of the quantum force, the wakefields can be generated with the
characteristic length scale $\lambda_{p}=2\pi(V_{g}^{2}-1)^{1/2}$ and in the
parameter regimes $B_{0}\sim10^{8}$ T, $\ n_{0}\sim10^{35}$ m$^{-3}$ as
described above$.$ Apart from the Fermi pressure, the quantum tunneling effect
associated with the Bohm potential introduces an additional higher-order
dispersion. The latter, in turn, introduces two characteristic length scales
$\lambda_{1}$ and $\lambda_{2}$ $\left(  <\lambda_{1}\right)  $ (quite
distinctive from the case with $H=0$) out of which only $\lambda_{1}$ favors
the wakefield excitation in the said regimes, since \ $\lambda_{2}$ becomes
smaller than the whistler wavelength $\lambda_{W}$ ($\gtrsim10^{-12}$ m), a
case not effective for the wakefield generation. However, multiple wakefield
excitation can be possible with two length scales $\lambda_{1}$ and
$\lambda_{2}$ in very strongly magnetized and superdense plasmas with
$B_{0}\gtrsim10^{9}$ T and $n_{0}\gtrsim10^{36}$ m$^{-3}.$ In these regimes
the CPF is strongly dominated by the SPF. \ The latter will then dominantly
accelerate the electrons by separating the electric charges and building up a
high electric field. \ Furthermore, such parameter regimes could be relevant
in the deeper layer (above the neutron-drip layer) of the crust of neutron
star in which electrons are degenerate implying that electrical and thermal
conductivity may be huge because the electrons can travel great distances
before interacting. However, \ the physics and the composition of neutron star
interiors are not yet fully understood. \

We ought to mention that in the regimes of very strong magnetic fields and
superdense plasmas as mentioned above, the nonrelativistic fluid model may no
longer be appropriate as in such cases the Fermi speed approaches (or can be
greater than) the speed of light in vacuum and whistler wavelength may become
smaller than the Compton wavelength of electrons. In this situation,
relativistic spin-quantum fluid model or kinetic approach could be well set.
By increasing the magnetic field strength or the ratio $\omega_{c}/\omega
_{p},$ we see that whistler wave can have smaller group dispersion. This could
be important for whistler waves to be a viable mechanism in magnetized dense
plasmas in order to gain higher energy for an accelerated particle from the
plasma wakefield.

Next, as an alternative mechanism instead of using lasers or particle beams as
drivers which have been used mostly for\ laboratory plasma wakefield
accelerations, the EM whistler waves should be of fundamental interest in
plasmas, in particular, quantum plasmas for astrophysical settings. However,
conclusive evidence needs further investigation in this area by considering,
e.g., full scale simulation of a relativistic fluid model or kinetic one
accounting for the quantum statistical and mechanical (tunneling) effects as
well as intrinsic spin of electrons. The latter might have a role in the
regimes previously considered to be as classical \cite{Spin-kinetic}. We
mention that our model can be well applicable for low or moderate density
magnetized plasmas as well where electrons may not be degenerate, spin effects
might not be so important. In this case one can consider, e.g., the isothermal
equation of state instead of the Fermi pressure law for electrons. \ Thus,
classical results can also be recovered by disregarding the term $\propto
\hbar$ for the spin contribution and the term $\propto\hbar^{2}$ for the
quantum tunneling effect.

To mention, neutron stars are known to be compact and carry intense surface
magnetic fields $\ $about $10^{9}$ T or more \cite{NeutronStar}. It has been
investigated that short gamma-ray bursts (GRBs) may arise (though, there are
still a lot of uncertainties about their origin) from collisions between a
black hole and a neutron star or between two neutron stars. However, when
neutron stars collide, the tremendous release of energy results into highly
relativistic out-bursting fireballs (jets) \cite{JETS}. The latter are most
likely in the form of a plasma and can have initial plasma density
$\gtrsim10^{32}$ m$^{-3}$. Such collisions of intense magnetic fields may
create strong magnetoshocks where whistler waves are embedded\textbf{.}

To conclude, we have presented a theoretical investigation for the possible
excitation of wakefields by self-consistent whistler wave field at nanoscale
in a magnetized spin quantum plasma. \ The whistler wave envelope is governed
by a NLS equation coupled to a driven Boussinesq-like equation for the lf
wakefield. \ The quantum force is shown to be responsible for the excitation
of multiple wakefields in strongly magnetized superdense plasmas in which SPF
strongly dominates over the CPF.\ Furthermore, the effect of quantum tunneling
is to enhance the wakefield amplitude, however it is reduced by the external
magnetic field.  Finally, the present investigation can be generalized to
multi-dimensional wakefield excitation in relativistic spin quantum plasmas.

\acknowledgments
APM acknowledges support from the Kempe Foundations, Sweden,
through Grant No. SMK-2647. MM was supported by the European Research Council
under Contract No. 204059-QPQV, and the Swedish Research Council under Contract No. 2007-4422.

\end{document}